\documentclass[10pt,twocolumn,twoside,submit]{IEEEtran}
\usepackage{color}
\usepackage{amsmath}
\usepackage{amssymb}
\usepackage{tabularx}
\usepackage{epsfig}
\usepackage{graphicx}
\usepackage[T1]{fontenc}
\usepackage{epstopdf}
\usepackage{cite}

\def\BibTeX{{\rm B\kern-.05em{\sc i\kern-.025em b}\kern-.08em
    T\kern-.1667em\lower.7ex\hbox{E}\kern-.125emX}}

\newtheorem{theorem}{Theorem}
\newtheorem{lemma}{Lemma}

\begin{document}
\bibliographystyle{jcn}

\title{
Amplify-and-Forward Two-Way Relaying System over Free-Space Optics Channels}
\author{Jaedon~Park~\IEEEmembership{Member,~IEEE,} Chan-Byoung~Chae~\IEEEmembership{Senior~Member,~IEEE,} and~Giwan~Yoon~\IEEEmembership{Member,~IEEE}
\thanks{This research was in part supported by the Agency for Defense Development (ADD) and the Basic Science Research Program through the National Research Foundation of Korea (NRF) funded by the Ministry of Education (Grant No. 2016R1D1A1B01007074).}
\thanks{J. Park is with the Agency for Defense Development (ADD), Daejeon, 34186, Korea. E-mail: jaedon@kaist.ac.kr.}
\thanks{G. Yoon is with the School of Electrical Engineering, Korea Advanced Institute of Science and Technology (KAIST), Daejeon, 34141, Korea. E-mail: gwyoon@kaist.ac.kr.}
\thanks{C.-B.~Chae is with the School of Integrated Technology, Yonsei University, 03722, Korea. E-mail: cbchae@yonsei.ac.kr.}
}
\markboth{}
{Park \lowercase{\textit{et al}}.: Two-Way Amplify-and-Forward Relaying System ...} \maketitle

\begin{abstract}
In this paper, we analyze the performance of a two-way subcarrier intensity-modulated (SIM) amplify-and-forward (AF) relaying system over free-space optics (FSO) communication channels. The analysis takes into consideration attenuations due to atmospheric turbulence, geometric spread and pointing errors at the same time. We derive, in generalized infinite power series expressions, the tight upper and lower bounds of the overall outage probability and average probability of errors of the system. The study finds that this two-way subcarrier intensity-modulated AF relaying system using a binary phase shift keying (BPSK) modulation could be used for practical applications in case of a weak turbulence regime in which the required SNR is about 30 dB to obtain the average bit error probability of $10^{-6}$. It is also noted that the pointing errors clearly degrade the performance of the two-way subcarrier intensity-modulated AF relaying system.
\end{abstract}

\begin{keywords}
Amplify-and-forward relay, atmospheric turbulence, average probability of error, free-space optics, overall outage probability, pointing errors, subcarrier intensity modulation, two-way relaying system, upper and lower bounds.
\end{keywords}

\vspace{10pt}
\section{\uppercase{Introduction}}
\label{sec:introd}
\PARstart{F}{ree}-space optics (FSO) systems are high-capacity and cost-effective communication techniques that are free of radio frequency spectrum regulations \cite{Nistazakis:09,Park:11,Uysal:06,Tsiftsis:09,park2015performance}. Because of this they have attracted enormous amounts of scholarly attention. Although an intensity modulation and direct detection (IM/DD) system using on/off keying (OOK) has been widely used due to its simplicity \cite{Nistazakis:09,Park:11,Uysal:06,Tsiftsis:09}, such a system is not appropriate for applications of amplify-and-forward (AF) relaying systems. Indeed, an IM/DD system requires an adaptive decision threshold that, in practice, is very difficult to implement. To circumvent this implementation difficulty, we consider in this paper a subcarrier intensity modulation (SIM) scheme. The scheme requires no adaptive decision threshold and ameliorates turbulence-induced irradiance fluctuation \cite{Popoola:09,Li:07,popoola2012,Peppas:10,samimi2010,song2012optical}. Therefore, the subcarrier intensity modulation scheme is suitable for AF relaying systems over FSO channels.

FSO systems are highly affected by attenuations caused by atmospheric turbulence and geometric spread and pointing errors \cite{sandalidis2008ber,sandalidis2009optical,farid2007outage,gappmair2011further,park2016impact}. As a result of variations in the refractive index, turbulence-induced fading, also known as scintillation, causes irradiance fluctuations in the received signals intensity \cite{sandalidis2008ber,sandalidis2009optical,farid2007outage,gappmair2011further}. Apart from the scintillation effects, pointing errors due to building sway also cause significant performance degradations in FSO systems \cite{sandalidis2008ber,sandalidis2009optical,farid2007outage,gappmair2011further}.

In addition to the performance degradation caused by atmospheric turbulence and pointing errors, FSO communication systems suffer a significant degradation in non-line-of-sight environments. Overcoming such problems is left up to relaying technologies \cite{Tsiftsis:06,safari2008relay,Lee:11,ansari2013impact,ansari2013SIECPC,ansari2013VTC,laneman2004cooperative,park2015outage}. Relaying systems are classified as either amplify-and-forward (AF) or decode-and-forward (DF) relays~\cite{laneman2004cooperative,Hasna:04,jaedonmilcom}. DF relaying systems, also called regenerative systems, decode the received signal fully and re-encode it before retransmitting it to another hop\cite{laneman2004cooperative,Hasna:04,jaedonmilcom}. The AF relaying systems, also called nonregenerative systems, just amplify the received signal and forward it to another hop with less complexity than the DF relaying systems \cite{laneman2004cooperative,Hasna:04,jaedonmilcom}. Since AF relaying systems do not decode and re-encode the received signal, they need less power and lower system complexity than do DF relaying systems \cite{laneman2004cooperative,Hasna:04,jaedonmilcom,Suraweera:09,chae_tsp_08,jcn_1, jcn_2}.

Recently, an enormous amount of research interest has been devoted to the two-way relaying system techniques for conventional RF applications \cite{upadhyay:11,guo2011symbol,jang2010,han2009performance,yang2011exact,ikki2012performance}. Such interest is  largely due to the technique's even more efficient signaling scheme, where two nodes bi-directionally communicate in just two phases via a half-duplex relay, resulting in the improvement of the spectral efficiency \cite{upadhyay:11,guo2011symbol,jang2010,han2009performance,yang2011exact,ikki2012performance}. Upadhyay \textit{et al.} studied the performance of a two-way opportunistic relaying system with analog network coding over Nakagami-m fading channels \cite{upadhyay:11}. Guo \textit{et al.} analyzed the overall outage probability as well as the symbol error probability of a two-way AF  relaying system over an exponential distribution \cite{guo2011symbol}. Jang \textit{et al.} have investigated the performance of a multiuser two-way relay channel \cite{jang2010}. Han \textit{et al.} have analyzed the tight upper and lower bounds of the average sum rate of a two-way AF relaying system with/without Alamouti's orthogonal space time block code (OSTBC) considering gamma distributions \cite{han2009performance}. Yang \textit{et al.} analyzed the performance of two-way AF relaying system in Nakagami-m fading channels \cite{yang2011exact}. Ikki \textit{et al.} have analyzed the performance of two-way AF relaying in the presence of co-channel interferences over Rayleigh fading channels \cite{ikki2012performance}. As is the case with a general relaying system, the two-way AF relaying system is more attractive in practice than the two-way DF relaying system due to its very simple processing at the relay terminal \cite{han2009performance}. Also, similarly to conventional AF relaying systems, the two-way AF relaying systems are able to lower  more effectively the system power consumption and complexity than are two-way DF relaying systems.

In the case of the two-way relaying systems for FSO applications, Tang \textit{et al.} proposed a network-coded cooperation relay scheme for optical DF two-way relay networks; they investigated the optimal bit decision algorithm for a receiving node over Gamma-Gamma fading distributions \cite{tang2011}. Puri \textit{et al.} investigated the performance of a two-way relay-assisted FSO system considering a DF protocol over log-normal distribution or gamma-gamma distribution\cite{Puri:13,puri2014outage}. They also proposed and analyzed relay selection protocols of two-way DF FSO relays assuming Gamma-Gamma distribution with pointing errors \cite{puri2015partial}. The same authors later analyzed two-way AF relay selection and derived the achievable-rates \cite{puri2015two} and outage probability and average error probability \cite{puri2015asymptotic} in closed-form expressions. Abu-Almaalie \textit{et al.} studied two-way DF FSO relay systems using SIM-BPSK \cite{abu2015physical}. 

It is difficult to gain an immediate insight from the analyzed performances of FSO systems \cite{Bayaki:09} since the error probabilities of FSO channels are usually expressed as a complex Meijer's G-function due to the modified Bessel function of the second kind \cite{Tsiftsis:06,Lee:11,tang2011}. Moreover, performance analysis is a  large challenge, especially in the case of AF relaying systems over a Gamma-Gamma fading distribution, because of the complex form of the Meijer's G-function. In a novel approach, \cite{Bayaki:09,Park:11} simplify the mathematical expressions by introducing a generalized infinite power series representation of the modified Bessel function of the second kind. With this representation, we can express the error probabilities of FSO systems as power series expansions, composed of only elementary and Gamma functions; this offers readers an mathematical insight into FSO systems.

In this paper, we analyze, for the first time, the performance of a two-way AF relaying system using the subcarrier intensity modulation scheme \cite{Popoola:09,Li:07,popoola2012,Peppas:10,samimi2010} over Gamma-Gamma fading environments with a simple generalized infinite power series expression \cite{Bayaki:09,Park:11}. We analyze FSO communication channels considering attenuations due to atmospheric turbulence and geometric spread and pointing errors \cite{sandalidis2008ber,sandalidis2009optical,farid2007outage,gappmair2011further}. The main contributions of this paper are as follows:

\begin{itemize}
	\item[\textbullet] \textit{Overall outage probability analysis of a two-way AF relaying system over FSO channels}: We derive the upper and lower bounds of the overall outage probability of a two-way AF relaying system over  Gamma-Gamma fading channels considering attenuations due to atmospheric turbulence and geometric spread and pointing errors.
\end{itemize}

\begin{itemize}
	\item[\textbullet] \textit{Average probability of error analysis of a two-way AF relaying system over FSO channels}: Based on the derived results of the overall outage probability upper and lower bounds, we further derive the average probability of errors of the two-way subcarrier intensity-modulated AF relaying system over the Gamma-Gamma fading channels considering attenuations due to atmospheric turbulence and geometric spread and pointing errors.
\end{itemize}

The rest of the paper is organized as follows: In Section~\ref{sec:System Model}, we discuss the system and channel model of a two-way subcarrier intensity-modulated AF relaying system over the Gamma-Gamma fading distributions. In Section \ref{sec:OOP analysis}, we derive, in generalized infinite power series expressions, the upper and lower bounds of the overall outage probability for the system under consideration. In Section \ref{sec:Avg Probability analysis}, we derive the average probability of errors corresponding to the upper and lower bounds of the overall outage probability given in Section \ref{sec:OOP analysis}. In Sections \ref{sec:Numerical results} and \ref{sec:Conclusion}, we present the numerical results and the conclusion. 

\begin{figure}[!t]
	\centering
	\includegraphics[width=3.4in]{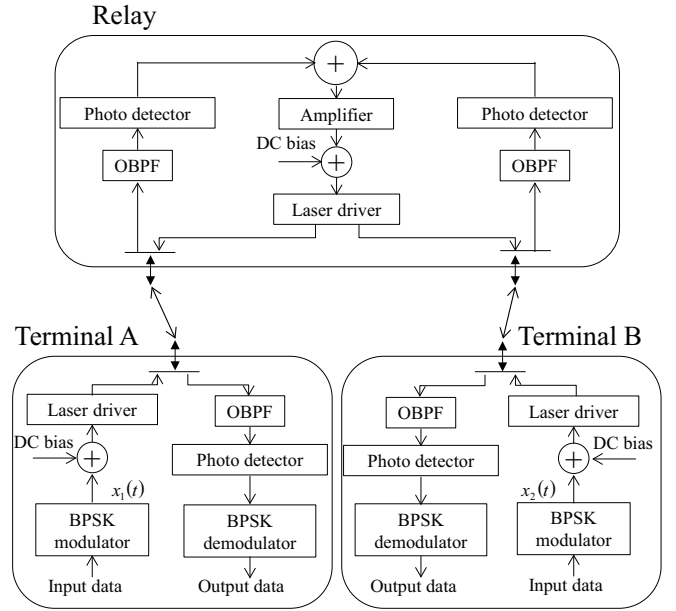}
	\caption{A two-way SIM AF relaying system using a BPSK modulation.}
	\label{fig_system}
\end{figure}

\vspace{10pt}
\section{System and channel model} \label{sec:System Model}
\subsection{System Model}
Consider a single-relay, two-way AF relaying system as described in Fig.~\ref{fig_system}, where the two source terminals, A and B, communicate along with a relay terminal R using the optical subcarrier intensity modulation scheme \cite{Popoola:09,Li:07,popoola2012,Peppas:10,samimi2010} with a binary phase shift keying (BPSK) modulation. For simplicity, throughout this paper, we assume a BPSK modulation. Other modulation schemes, however, can also be applicable. The channels are assumed to be stationary with independent and identically distributed (i.i.d.) intensity fading statistics. The channel state information (CSI) is also assumed to be available at the receiver.

At Terminal A, the source data is modulated onto the RF subcarrier signal $x_{1}(t)$ using a BPSK modulation.  The RF signal is added with a bias signal to drive the laser diode on the positive values. At Relay R, the received optical signal radiated from Terminal A passes through an optical band pass filter (OBPF) to reject the background radiation noise. Finally, the photodetector generates the photocurrent $y_{R1}(t)$ proportional to $x_{1}(t)$. Similarly, the photodetector generates the photocurrent $y_{R2}(t)$ proportional to $x_{2}(t)$, the RF subcarrier signal radiated from the source, Terminal B. After the two signals $y_{R1}(t)$ and $y_{R2}(t)$  are added, the photocurrent can be given by
\begin{equation} \label{eq:eq_5a}
	y_{R}(t)=\eta I_{1} (1 + \xi x_{1}(t)) + \eta I_{2} (1 + \xi x_{2}(t)) + n_{R}(t). \nonumber
\end{equation}
where $\eta$  is the photodetector responsiveness; $I_{1}$  is the irradiance from Terminal A to Relay R; $I_{2}$  is the irradiance from Terminal B to Relay R; $\xi$ is a constant to fultill  $\lvert \xi x_{1}(t) \rvert \leq 1$, and $\lvert \xi x_{2}(t) \rvert \leq 1 $; $n_{R}(t) \sim\mathcal{N}(0,\sigma^{2})$  is an additive white Gaussian noise (AWGN) mainly due to thermal and/or background noise \cite{Popoola:09,samimi2010,song2012optical}. The channel state is considered to be the product of two random factors, i.e., $I_i = I_{ia} I_{ip}$ where $I_{ia}$ is the attenuation due to atmospheric turbulence, which is modeled as Gamma-Gamma fading distribution\cite{Popoola:09,Uysal:06,Peppas:10,Nistazakis:09,Bayaki:09,Park:11}, and $I_{ip}$ is the attenuation due to geometric spread and pointing errors \cite{sandalidis2008ber,sandalidis2009optical,farid2007outage,gappmair2011further,ansari2013impact,ansari2013SIECPC,ansari2013VTC}.
It is noted that the channel model of random variable $I_{i}$, i = 1, 2, is described in detail in the following section.

After the DC components of the received photocurrent $y_{R}(t)$ are filtered out, the photocurrent can be given by
\begin{equation} \label{eq:eq_5}
	y_{R}(t)=\eta I_{1} \xi x_{1}(t) + \eta I_{2} \xi x_{2}(t) + n_{R}(t).
\end{equation}
The relay amplifies the electrical signal by the gain, $G_{R}$, which is given by
\begin{equation} \label{eq:eq_6}
	G_{R}= \sqrt{\frac{P_{R}}{P_{S}(\eta I_{1} \xi)^2 + P_{S}(\eta I_{2} \xi)^2 + \sigma^2}},
\end{equation}
to meet the average transmit power constraints \cite{guo2011symbol}. The amplified electrical signal is added with a bias signal to drive the laser diode on the positive values. Finally, the relay retransmits the optical signal to both terminals, A and B. Here, $P_{R}$  and $P_{S}$  represent the subcarrier signal powers of the relay and the terminal, respectively.
At Terminal A, after passing through an optical band pass filter, the received photocurrent is given by
\begin{equation} \label{eq:eq_7}
	y_{BRA}(t)=\eta I_{1} (1+ \xi G_{R}y_{R}(t)) + n_{A}(t)
\end{equation}
where $n_{A}(t)$  is the AWGN with $\sim\mathcal{N}(0,\sigma^2)$.
Substituting ($\ref{eq:eq_5}$) for ($\ref{eq:eq_7}$) and filtering out the DC component and subtracting the self-interference parts \cite{upadhyay:11,guo2011symbol,jang2010,tang2011}, the received photocurrent can be rewritten as
\begin{equation} \label{eq:eq_8}
	y_{BRA}(t)=G_{R} \eta I_{1} \xi \eta I_{2}\xi x_{2}(t) + G_{R} \eta I_{1} \xi n_{R}(t) + n_{A}(t).
\end{equation}

Substituting ($\ref{eq:eq_6}$) into ($\ref{eq:eq_8}$), without any loss of generality, if we assume both  $P_{S} = P_{R} = P_{0}$ and $P_{0} \gg \sigma^2$, the received SNR at Terminal A can be expressed as
\begin{equation} \label{eq:eq_9}
	\Gamma_{BRA} = \frac{(I_{1}I_{2})^2}{2I_{1}^2 + I_{2}^2} \eta^2  \xi^2 \gamma_{0}, \nonumber
\end{equation}
where  $\gamma_{0} = P_{0} / \sigma^2$ is the SNR in the absence of turbulence and pointing errors. Similarly, the received SNR at Terminal B can be expressed as
\begin{equation} \label{eq:eq_10}
	\Gamma_{ARB} = \frac{(I_{1}I_{2})^2}{I_{1}^2 + 2I_{2}^2} \eta^2  \xi^2 \gamma_{0}. \nonumber
\end{equation}

\subsection{Turbulence and Misalignment Fading Model}
It is assumed that the random variable $I_{ia}$, i = 1, 2, follows a Gamma-Gamma distribution, and using the generalized power series representation method, the probability density function (PDF) and the cumulative distribution function (CDF) of the random variable  $I_{ia}$ can be expressed as  \cite{Bayaki:09,Park:11}
\begin{equation} \label{eq:eq_pdfBayaki}
	f_{I_{ia}}(I_{ia})=\lim_{J \to \infty} \sum_{j=0}^J \left(a_{j}(\alpha,\beta)I_{ia}^{j+\beta-1} + a_{j}(\beta,\alpha)I_{ia}^{j+\alpha-1}\right),
\end{equation}
\begin{equation} \label{eq:eq_cdfBayaki}
	F_{I_{ia}}(I_{ia})=\lim_{J \to \infty} \sum_{j=0}^J \left(\frac{a_{j}(\alpha,\beta)}{j+\beta} I_{ia}^{j+\beta} + \frac{a_{j}(\beta,\alpha)}{j+\alpha} I_{ia}^{j+\alpha}\right),
\end{equation}
where $a_{j}(\alpha,\beta) = \frac{\pi(\alpha\beta)^{j+\beta}}{\sin[\pi(\alpha-\beta)]\Gamma(\alpha)\Gamma(\beta)\Gamma(j-\alpha+\beta+1)j!}$, and $\alpha$  and   $\beta$ are the atmospheric turbulence parameters \cite{Popoola:09,Uysal:06,Peppas:10,Park:11,Bayaki:09}
\begin{equation} \label{eq:eq_alpha}
	\alpha = \left[\text{exp}{\left(\frac{0.49\chi^2}{(1+0.18d^2 + 0.56\chi^{12/5})^{7/6}}\right)} - 1 \right]^{-1},\nonumber
\end{equation}
\begin{equation} \label{eq:eq_beta}
	\beta = \left[\text{exp}{\left(\frac{0.51\chi^2(1+0.69\chi^{12/5})^{-5/6}}{(1+0.9d^2 + 0.62d^2\chi^{12/5})^{5/6}}\right)} - 1 \right]^{-1},\nonumber
\end{equation}
where $\chi^2 = 0.5 C_n^2 k^{7/6} z^{11/6}$ and $d = (k D^2 / 4z)^{1/2}$. Here $k=2\pi / \lambda$ is the optical wave number, $\lambda$ is the wavelength, $D$ is the diameter of the receiver's collecting lens aperture, $z$ is the link distance in meters and $C_n^2$ is the altitude-dependent index of refraction structure (see \cite{Popoola:09,Uysal:06,Peppas:10,Bayaki:09} for more details).

The PDF of $I_{ip}$, by considering a circular detection aperture of radius $r$ and a Gausssian beam, is given by \cite{sandalidis2008ber,sandalidis2009optical,farid2007outage,gappmair2011further}
\begin{equation} \label{eq:eq_pdfPointError}
	f_{I_{ip}}(I_{ip})=\frac{\gamma^2}{A_0^{\gamma^2}}I_{ip}^{\gamma^2-1}, ~~~0 \leq I_{ip} \leq A_0
\end{equation}
where $\gamma=w_{z_{eq}}/2\sigma_s$ is the ratio between the equivalent beam radius at the receiver and the pointing error displacement standard deviation (jitter) at the receiver \cite{sandalidis2008ber,sandalidis2009optical,farid2007outage,gappmair2011further}. The parameter $w_{z_{eq}}$ can be calculated using the relation $v=\sqrt{\pi}r/\sqrt{2}w_z$, $A_0=[\text{erf}(v)]^2$ and $w_{z_{eq}}^2=w_z^2 \sqrt{\pi}\text{erf}(v)/2v\exp{(-v^2)}$ where \text{erf}($\cdot$) is the error function and $w_z$ is the beam waist (radius calculated at $e^{-2}$) at distance $z$ \cite{sandalidis2008ber,sandalidis2009optical,farid2007outage,gappmair2011further}.
The combined PDF of $I_i = I_{ia} I_{ip}$ is given by \cite{sandalidis2008ber,sandalidis2009optical,farid2007outage,gappmair2011further}
\begin{equation} \label{eq:eq_pdfcombined}
	f_{I_{i}}(I_{i})=\int f_{I_i|I_{ia}} (I_i|I_{ia}) f_{I_{ia}}(I_{ia})dI_{ia},
\end{equation}
where $f_{I_i|I_{ia}} (I_i|I_{ia})$ is the conditional probability given $I_{ia}$ state and is expressed by
\begin{multline} \label{eq:eq_pdfhha}
	f_{I_i|I_{ia}} (I_i|I_{ia}) = \frac{1}{I_{ia}} f_{I_{ip}} \left(\frac{I_i}{I_{ia}}\right) \\
	= \frac{\gamma^2}{A_0^{\gamma^2}I_{ia}} \left(\frac{I_i}{I_{ia}}\right) ^{\gamma^2-1}, ~~~0 \leq I_{i} \leq A_0 I_{ia}.
\end{multline}

Substituting ($\ref{eq:eq_pdfBayaki}$) and ($\ref{eq:eq_pdfhha}$), ($\ref{eq:eq_pdfcombined}$) can be rewritten as
\begin{multline} \label{eq:eq_pdfcombinedInt}
	f_{I_{i}}(I_{i})= \frac{\gamma^2}{A_0^{\gamma^2}} I_i^{\gamma^2-1}
	\int_{I_i/A_0}^{\infty} \times \\
	\lim_{J \to \infty} \sum_{j=0}^J \left(a_{j}(\alpha,\beta)I_{ia}^{j+\beta-1-\gamma^2} + a_{j}(\beta,\alpha)I_{ia}^{j+\alpha-1-\gamma^2}\right)dI_{ia}.
\end{multline}

By calculating the definite integration, ($\ref{eq:eq_pdfcombinedInt}$) is derived to
\begin{multline} \label{eq:eq_pdfcombinedfinal}
	f_{I_{i}}(I_{i})= \frac{\gamma^2}{A_0^{\gamma^2}} I_i^{\gamma^2-1}
	\lim_{J \to \infty} \times \\
	\sum_{j=0}^J \left(\frac{a_{j}(\alpha,\beta) (I_{i}/A_0)^{j+\beta-\gamma^2} }{\gamma^2-j-\beta} + \frac{a_{j}(\beta,\alpha)(I_{i}/A_0)^{j+\alpha-\gamma^2}}{\gamma^2-j-\alpha}\right).
\end{multline}

By integrating the PDF in ($\ref{eq:eq_pdfcombinedfinal}$), the CDF can be obtained as
\begin{align} \label{eq:eq_cdfcombinedfinal}F_{I_{i}}(I_{i})= \frac{\gamma^2}{A_0} & \lim_{J \to \infty} \sum_{j=0}^J  \Bigg(\frac{a_{j}(\alpha,\beta) I_{i}^{j+\beta} }{(j+\beta)(\gamma^2-j-\beta)A_0^{j+\beta-1}} \nonumber \\
	&   + \frac{a_{j}(\beta,\alpha) I_{i}^{j+\alpha}} {(j+\alpha)(\gamma^2-j-\alpha)A_0^{j+\alpha-1}}\Bigg).
\end{align}
It is easily confirmed that when $\gamma^2$ goes to $\infty$ (for the non-pointing errors case), ($\ref{eq:eq_pdfcombinedfinal}$) and ($\ref{eq:eq_cdfcombinedfinal}$) converge to ($\ref{eq:eq_pdfBayaki}$) and ($\ref{eq:eq_cdfBayaki}$), respectively.

\section{Overall outage probability analysis} \label{sec:OOP analysis}
Before we analyze the overall outage probability, we first study the integral of a rectangular region as described in~Fig.~2.
\begin{lemma}
	Integral of a rectangular region for $1/I_{1}^2$, $1/I_{2}^2$.
\end{lemma}
The rectangular integral region can be expressed as
\begin{equation} \label{eq:eq_14}
	\Psi(a,c) = \{1 - F_{I_{2}}(1/\sqrt{a})\}\{1 - F_{I_{1}}(1/\sqrt{c}) \}.
\end{equation}

\begin{figure}[!t]
	\centering
	\includegraphics[width=1.2in]{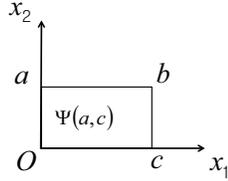}
	\caption{Rectangular integral region for $X_{1}$ and $X_{2}$.}
	\label{fig_lemma}
\end{figure}

\begin{proof}
	For a random variable  $I_{i}$, $i$ = 1, 2, we define a new random variable  $X_{i}=1/I_{i}^2$. The integral of a rectangular region composed of  $X_{1}$ and $X_{2}$, as shown in Fig. 2, can be written as
	\begin{multline} \label{eq:eq_11}
		\Psi(a,c) = \int_{0}^{c} \Pr[X_{2} < a | x_{1}]f_{X_{1}}(x_{1})dx_{1} \\
		=\int_{0}^{c}\int_{0}^{a}  f_{X_{2}}(x_{2})dx_{2} f_{X_{1}}(x_{1})dx_{1}.~~~~~~~~~~~~~~~~\nonumber
	\end{multline}
	The two random variables, $I_{1}$ and $I_{2}$, are assumed to be independent but not necessarily identically distributed Gamma-Gamma fading. Hence, the above rectangular integral region can be rewritten as
	\begin{equation} \label{eq:eq_12}
		\Psi(a,c) = F_{X_{2}}(a) F_{X_{1}}(c).
	\end{equation}
	Since  $X_{i}=1/I_{i}^2$,  $i$ = 1, 2, the CDF of  $X_{i}$ can be given by
	\begin{multline} \label{eq:eq_13}
		F_{X_{i}}(x) = \Pr(X_{i} < x) = \Pr(I_{i} > I) = 1-\Pr(I_{i} < I) \\
		= 1 - F_{I_{i}}(I) = 1 - F_{I_{i}}(1/\sqrt{x}). ~~~~~~~~~~~~~~
	\end{multline}
	Substituting ($\ref{eq:eq_13}$) into ($\ref{eq:eq_12}$), ($\ref{eq:eq_14}$) is obtained.
\end{proof}

For the two-way subcarrier intensity-modulated AF relaying system, the overall outage probability is, then, defined as
\begin{multline} \label{eq:eq_15}
	P_\text{out}=\Pr\left[\min\left(\frac{(I_{1}I_{2})^2}{2I_{1}^2 + I_{2}^2} \eta^2  \xi^2 \gamma_{0}, \frac{(I_{1}I_{2})^2}{I_{1}^2 + 2I_{2}^2} \eta^2  \xi^2 \gamma_{0}\right)<\Gamma_\text{th}\right]\\
	~~~~~~~~~~~=1-\Pr\left[\frac{(I_{1}I_{2})^2}{2I_{1}^2 + I_{2}^2} > \frac{\Gamma_\text{th}}{\eta^2  \xi^2 \gamma_{0}}, \frac{(I_{1}I_{2})^2}{I_{1}^2 + 2I_{2}^2} > \frac{\Gamma_\text{th}}{\eta^2  \xi^2 \gamma_{0}}\right]
\end{multline}
where, $\Gamma_\text{th}$ is the threshold SNR.
Let  $X_{i}=1/I_{i}^2$,  $i$ = 1, 2, the outage probability can be rewritten as
\begin{equation} \label{eq:eq_16}
	P_\text{out}=1-\Phi(\gamma_{0},\Gamma_\text{th})
\end{equation}
where,
\begin{multline} \label{eq:eq_17}
	\Phi(\gamma_{0},\Gamma_\text{th})\\
	=\Pr\left[X_{2} < -\frac{X_{1}}{2} + \frac{\eta^2  \xi^2 \gamma_{0}}{2\Gamma_\text{th}}, X_{2} < -2X_{1} + \frac{\eta^2  \xi^2 \gamma_{0}}{\Gamma_\text{th}}\right].
\end{multline}

The upper probability $\Phi(\gamma_{0},\Gamma_\text{th})$ is very difficult to derive in a closed-form \cite{guo2011symbol}. As described in Fig. 3, the integral region for the calculation of  $\Phi(\gamma_{0},\Gamma_\text{th})$ is $OABC$. Here, we analyze the upper and lower bounds instead of the exact form.

\begin{figure}[!t]
	\centering
	\includegraphics[width=2.5in]{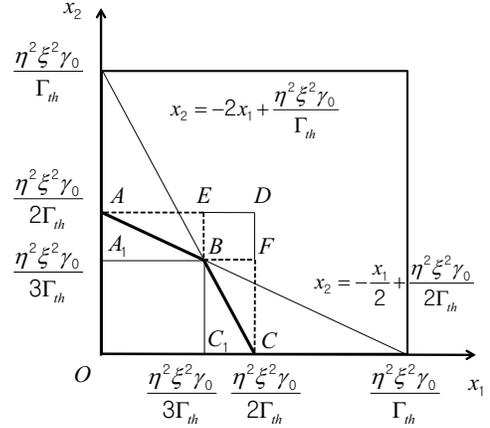}
	\caption{Integral region of $OABC$ for $\Phi(\gamma_{0},\Gamma_\text{th})$.}
	\label{fig_intRegion}
\end{figure}

\subsection{The upper bound of the overall outage probability}
\begin{theorem}
	The upper bound of the overall outage probability is given by
\end{theorem}
\begin{equation} \label{eq:eq_21}
	P_\text{out} < P_\text{out,U} = 1 - \Psi\left(\frac{\eta^2  \xi^2 \gamma_{0}}{3\Gamma_\text{th}},\frac{\eta^2  \xi^2 \gamma_{0}}{3\Gamma_\text{th}}\right).
\end{equation}

\begin{proof}
	The integral region $OABC$ is lower bounded by $OA_{1}BC_{1}$. Here, the integral region, $OA_{1}BC_{1}$, can be obtained by \emph{Lemma 1} as follows:
	\begin{equation} \label{eq:eq_19}
		OA_{1}BC_{1} = \Psi\left(\frac{\eta^2  \xi^2 \gamma_{0}}{3\Gamma_\text{th}},\frac{\eta^2  \xi^2 \gamma_{0}}{3\Gamma_\text{th}}\right).
	\end{equation}
	Thus, the integral region of $OABC$, $\Phi(\gamma_{0},\Gamma_\text{th})$, is lower-bounded by
	\begin{equation} \label{eq:eq_20}
		\Phi(\gamma_{0},\Gamma_\text{th}) > \Psi\left(\frac{\eta^2  \xi^2 \gamma_{0}}{3\Gamma_\text{th}},\frac{\eta^2  \xi^2 \gamma_{0}}{3\Gamma_\text{th}}\right).
	\end{equation}
	Substituting ($\ref{eq:eq_20}$) into ($\ref{eq:eq_16}$), ($\ref{eq:eq_21}$) is obtained.
\end{proof}

Also, the CDF, $F_{\Gamma}^{U}(x)$, of the $\min\left(\frac{(I_{1}I_{2})^2}{2I_{1}^2 + I_{2}^2} \eta^2  \xi^2 \gamma_{0}, \frac{(I_{1}I_{2})^2}{I_{1}^2 + 2I_{2}^2} \eta^2  \xi^2 \gamma_{0}\right)$ is approximated from the outage probability upper bound in ($\ref{eq:eq_21}$) by the variable change from $\Gamma_\text{th}$ to $x$ as
\begin{equation} \label{eq:eq_22}
	F_{\Gamma}^{U}(x) = 1 - \Psi\left(\frac{\eta^2  \xi^2 \gamma_{0}}{3x},\frac{\eta^2  \xi^2 \gamma_{0}}{3x}\right).
\end{equation}

\subsection{The lower bound of the overall outage probability}
\begin{theorem}
	The lower bound of the overall outage probability is given by
\end{theorem}
\begin{multline} \label{eq:eq_26}
	P_\text{out} > P_\text{out,L} = 1 - \Bigg\{\Psi\left(\frac{\eta^2  \xi^2 \gamma_{0}}{2\Gamma_\text{th}},\frac{\eta^2  \xi^2 \gamma_{0}}{3\Gamma_\text{th}}\right) \\
	+ \Psi\left(\frac{\eta^2  \xi^2 \gamma_{0}}{3\Gamma_\text{th}},\frac{\eta^2  \xi^2 \gamma_{0}}{2\Gamma_\text{th}}\right)
	- \Psi\left(\frac{\eta^2  \xi^2 \gamma_{0}}{3\Gamma_\text{th}},\frac{\eta^2  \xi^2 \gamma_{0}}{3\Gamma_\text{th}}\right)\Bigg\}.
\end{multline}

\begin{proof}
	The integral region $OABC$ is upper bounded by $OAEBFC$. The integral region of $OAEBFC$ can be expressed as
	\begin{equation} \label{eq:eq_23}
		OAEBFC = OAEC_{1} + OA_{1}FC - OA_{1}BC_{1}.
	\end{equation}
	Here, each integral of all regions, $OAEC_{1}$, $OA_{1}FC$, and $OA_{1}BC_{1}$, can be obtained by \emph{Lemma 1} as follows:
	\begin{multline} \label{eq:eq_24}
		OAEC_{1} : \Psi\left(\frac{\eta^2  \xi^2 \gamma_{0}}{2\Gamma_\text{th}},\frac{\eta^2  \xi^2 \gamma_{0}}{3\Gamma_\text{th}}\right), \\
		OA_{1}FC  : \Psi\left(\frac{\eta^2  \xi^2 \gamma_{0}}{3\Gamma_\text{th}},\frac{\eta^2  \xi^2 \gamma_{0}}{2\Gamma_\text{th}}\right), ~~~~~~~~~~~~~~~~~~~~~~~~~~~~\\
		OA_{1}BC_{1} : \Psi\left(\frac{\eta^2  \xi^2 \gamma_{0}}{3\Gamma_\text{th}},\frac{\eta^2  \xi^2 \gamma_{0}}{3\Gamma_\text{th}}\right).~~~~~~~~~~~~~~~~~~~~~
	\end{multline}
	Thus, the integral region of $OABC$, $\Phi(\gamma_{0},\Gamma_\text{th})$, is upper-bounded by
	\begin{multline} \label{eq:eq_25}
		\Phi(\gamma_{0},\Gamma_\text{th}) < \Psi\left(\frac{\eta^2  \xi^2 \gamma_{0}}{2\Gamma_\text{th}},\frac{\eta^2  \xi^2 \gamma_{0}}{3\Gamma_\text{th}}\right) \\
		+ \Psi\left(\frac{\eta^2  \xi^2 \gamma_{0}}{3\Gamma_\text{th}},\frac{\eta^2  \xi^2 \gamma_{0}}{2\Gamma_\text{th}}\right)
		- \Psi\left(\frac{\eta^2  \xi^2 \gamma_{0}}{3\Gamma_\text{th}},\frac{\eta^2  \xi^2 \gamma_{0}}{3\Gamma_\text{th}}\right).
	\end{multline}
	Substituting ($\ref{eq:eq_25}$) into ($\ref{eq:eq_16}$), ($\ref{eq:eq_26}$) is obtained.
\end{proof}

Also, the CDF, $F_{\Gamma}^{L}(x)$, of the $\min\left(\frac{(I_{1}I_{2})^2}{2I_{1}^2 + I_{2}^2} \eta^2 \xi^2 \gamma_{0}, \frac{(I_{1}I_{2})^2}{I_{1}^2 + 2I_{2}^2} \eta^2 \xi^2 \gamma_{0}\right)$ is approximated from the outage probability lower bound in ($\ref{eq:eq_26}$) by the variable change from $\Gamma_\text{th}$ to $x$ as
\begin{multline} \label{eq:eq_27}
	F_{\Gamma}^{L}(x) = 1 - \Bigg\{\Psi\left(\frac{\eta^2 \xi^2 \gamma_{0}}{2x},\frac{\eta^2 \xi^2 \gamma_{0}}{3x}\right) \\
	+ \Psi\left(\frac{\eta^2 \xi^2 \gamma_{0}}{3x},\frac{\eta^2 \xi^2 \gamma_{0}}{2x}\right)
	- \Psi\left(\frac{\eta^2 \xi^2 \gamma_{0}}{3x},\frac{\eta^2 \xi^2 \gamma_{0}}{3x}\right)\Bigg\}.
\end{multline}

\section{Average Probability of Error Analysis} \label{sec:Avg Probability analysis}
In this section, the average probability of error is derived for the two-way subcarrier intensity-modulated AF relaying system. If we let $P(e|\Gamma)$ denote the conditional error probability in an AWGN channel, the average probability of error can be expressed as

\begin{equation} \label{eq:eq_28}
	P_{e}=\int_{0}^{\infty}P(e|\Gamma)f_{\Gamma}(\Gamma)d\Gamma
\end{equation}
where, the conditional error probability can be given by \cite{Simon:00}

\begin{equation} \label{eq:eq_29}
	P(e|\Gamma)=Q\left(\sqrt{\delta\Gamma}\right)
\end{equation}
where, $\delta$ is 2 for BPSK modulation.
Substituting ($\ref{eq:eq_29}$) to ($\ref{eq:eq_28}$), ($\ref{eq:eq_28}$) can be rewritten as \cite{Suraweera:09}

\begin{equation} \label{eq:eq_30}
	P_{e} = \frac{1}{\sqrt{2\pi}} \int_{0}^{\infty}F_{\Gamma}\left(\frac{t^2}{\delta}\right)e^{-\frac{t^2}{2}}dt
\end{equation}
where, $F_{\Gamma}(\Gamma)$ is the CDF of the random variable $\Gamma$.
After the variable change of $x=t^2$, the average probability of error can be given by \cite{jaedonmilcom}

\begin{equation} \label{eq:eq_31}
	P_{e}=\frac{1}{\sqrt{2\pi}} \int_{0}^{\infty}F_{\Gamma}\left(\frac{x}{\delta}\right)e^{-\frac{x}{2}}\frac{dx}{2\sqrt{x}}.
\end{equation}

\begin{figure}[!t]
	\centering
	\includegraphics[width=3.5in]{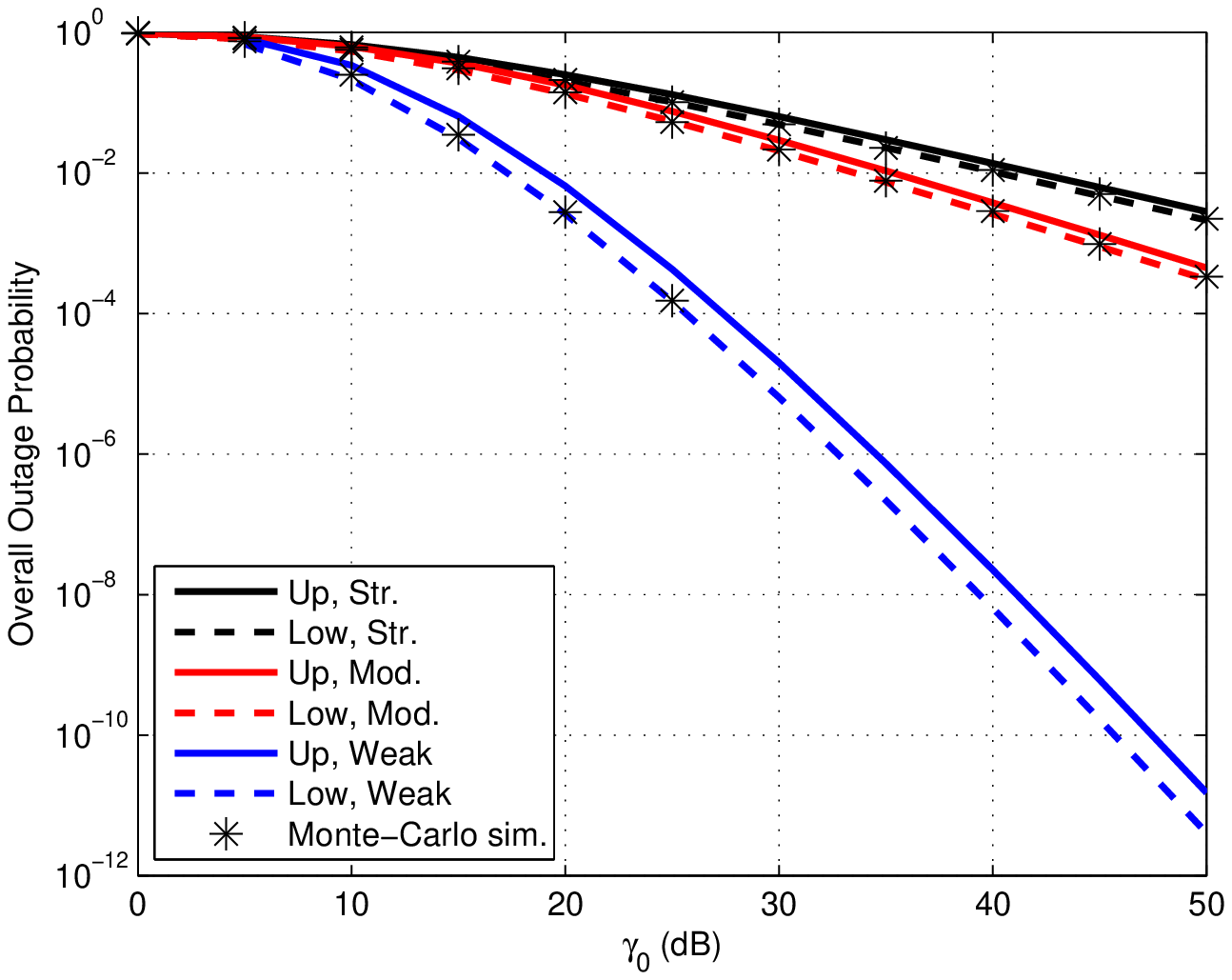}
	\caption{Overall outage probability of a two-way SIM AF relaying system. $J$ is set to 100. $\Gamma_\text{th}$ = 0 dB.}
	\label{fig_outage_0dB}
\end{figure}

\begin{theorem}
	If we substitute ($\ref{eq:eq_22}$) and ($\ref{eq:eq_27}$) into ($\ref{eq:eq_31}$), we obtain the average probability of errors corresponding to the upper and lower bounds of the overall outage probability as
\end{theorem}
\begin{equation} \label{eq:eq_35_1}
	P_{U}(e)=\frac{1}{2\sqrt{2\pi}} \Bigg\{\sqrt{2\pi} - A\left(\frac{\eta^2 \xi^2 \gamma_{0} \delta}{3},\frac{\eta^2 \xi^2 \gamma_{0} \delta}{3}\right)\Bigg\},
\end{equation}
\begin{multline} \label{eq:eq_36_1}
	P_{L}(e)=\frac{1}{2\sqrt{2\pi}} \Bigg\{\sqrt{2\pi} - A\left(\frac{\eta^2 \xi^2 \gamma_{0}\delta}{2},\frac{\eta^2 \xi^2 \gamma_{0}\delta}{3}\right) \\
	- A\left(\frac{\eta^2 \xi^2 \gamma_{0}\delta}{3},\frac{\eta^2 \xi^2 \gamma_{0}\delta}{2}\right) + A\left(\frac{\eta^2 \xi^2 \gamma_{0}\delta}{3},\frac{\eta^2 \xi^2 \gamma_{0}\delta}{3}\right)  \Bigg\},
\end{multline}
where,
\begin{equation} \label{eq:eq_Auv}
	A(u,v) = \sqrt{2\pi} - A_{1}^{J}(u) - A_{2}^{J}(v) + A_{3}^{J}(u,v).
\end{equation}
Here, $A_{1}^{J}(u)$, $A_{2}^{J}(v)$ and $A_{3}^{J}(u,v)$ are given by
\begin{align}
	\begin{split} \label{eq:eq_A1final}
		A_{1}^{J}(u) =&  \frac{\gamma^2}{A_0} \lim_{J \to \infty} \sum_{j=0}^J \frac{a_{j}(\alpha,\beta)u^{-\frac{j+\beta}{2}}}{(j+\beta)(\gamma^2-j-\beta)A_0^{j+\beta-1}} \\
		&\times 2^{\frac{j+\beta+1}{2}} \Gamma\left(\frac{j+\beta+1}{2}\right)\\
		&         + \frac{\gamma^2}{A_0} \lim_{J \to \infty} \sum_{j=0}^J \frac{a_{j}(\beta,\alpha)u^{-\frac{j+\alpha}{2}}}{(j+\alpha)(\gamma^2-j-\alpha)A_0^{j+\alpha-1}} \\
		&\times 2^{\frac{j+\alpha+1}{2}} \Gamma\left(\frac{j+\alpha+1}{2}\right),
	\end{split}
\end{align}

\begin{align}
	\begin{split} \label{eq:eq_A2final}
		A_{2}^{J}(v) =& \frac{\gamma^2}{A_0} \lim_{J \to \infty} \sum_{j=0}^J \frac{a_{j}(\alpha,\beta)v^{-\frac{j+\beta}{2}}}{(j+\beta)(\gamma^2-j-\beta)A_0^{j+\beta-1}} \\
		&\times 2^{\frac{j+\beta+1}{2}} \Gamma\left(\frac{j+\beta+1}{2}\right)\\
		&+ \frac{\gamma^2}{A_0} \lim_{J \to \infty} \sum_{j=0}^J \frac{a_{j}(\beta,\alpha)v^{-\frac{j+\alpha}{2}}}{(j+\alpha)(\gamma^2-j-\alpha)A_0^{j+\alpha-1}} \\
		&\times 2^{\frac{j+\alpha+1}{2}} \Gamma\left(\frac{j+\alpha+1}{2}\right),
	\end{split}
\end{align}

\begin{multline} \label{eq:eq_A3final}
	A_{3}^{J}(u,v)\\
	= \left(\frac{\gamma^2}{A_0}\right)^2 \lim_{J \to \infty} \sum_{j=0}^J c_{j}^{(1p)}(u,v) 2^{\frac{j+2\beta+1}{2}} \Gamma\left(\frac{j+2\beta+1}{2}\right)\\
	+ \left(\frac{\gamma^2}{A_0}\right)^2 \lim_{J \to \infty} \sum_{j=0}^J c_{j}^{(2p)}(u,v) 2^{\frac{j+\alpha+\beta+1}{2}} \Gamma\left(\frac{j+\alpha+\beta+1}{2}\right)\\
	+ \left(\frac{\gamma^2}{A_0}\right)^2 \lim_{J \to \infty} \sum_{j=0}^J c_{j}^{(3p)}(u,v) 2^{\frac{j+\alpha+\beta+1}{2}} \Gamma\left(\frac{j+\alpha+\beta+1}{2}\right)\\
	+ \left(\frac{\gamma^2}{A_0}\right)^2 \lim_{J \to \infty} \sum_{j=0}^J c_{j}^{(4p)}(u,v) 2^{\frac{j+2\alpha+1}{2}} \Gamma\left(\frac{j+2\alpha+1}{2}\right).
\end{multline}

\begin{proof}
	See Appendix A.
\end{proof}

\section{Numerical results} \label{sec:Numerical results}
In this section, we present the numerical results of the derived overall outage probability's upper and lower bounds and average probability of errors corresponding to the bounds of the overall outage probability. A two-way subcarrier intensity-modulated AF relaying system was considered, using a BPSK modulation over the FSO channel. The FSO channel was modeled as a Gamma-Gamma fading distribution with the atmospheric turbulence parameters of $\alpha$ = 4.2 and $\beta$  = 1.4 for a strong turbulence regime, parameters of $\alpha$  = 4.0 and   $\beta$ = 1.9 for a moderate turbulence regime, and $\alpha$  = 8.5  and $\beta$  = 6.7, for a weak turbulence regime \cite{Lee:11,Popoola:09,Park:11,Peppas:10}. For pointing errors, the normalized jitter of $\sigma_s/r = 0.02 \sim 0.4$ and the normalized beamwidth of $w_z/r = 5, 10, 15$ were considered. For numerical evaluations, the parameter $J$ used in the infinite power series was truncated to 100.

Figs. 4, 5, and 6 show the overall outage probability performances of the two-way subcarrier intensity-modulated AF relaying system with respect to the SNR in the absence of turbulence and pointing errors for the threshold SNRs of 0~dB, 5~dB, and 10~dB, respectively. Here, we do not consider pointing errors ($\gamma^2$ is set to $\infty$ with $A_0 = 1$).

\begin{figure}[!t]
	\centering
	\includegraphics[width=3.5in]{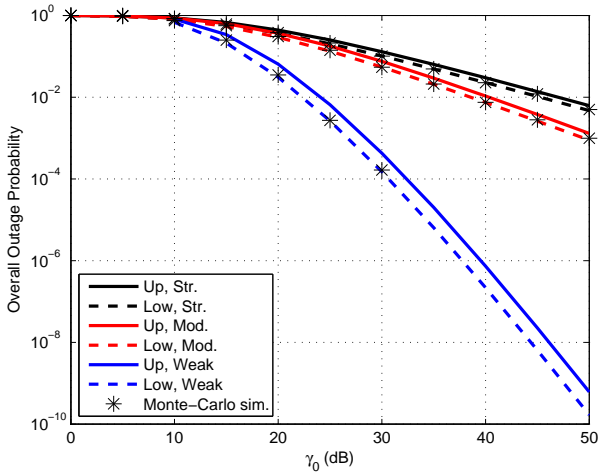}
	\caption{Overall outage probability of a two-way SIM AF relaying system. $J$ is set to 100. $\Gamma_\text{th}$ = 5 dB.}
	\label{fig_outage_5dB}
\end{figure}

\begin{figure}[!t]
	\centering
	\includegraphics[width=3.5in]{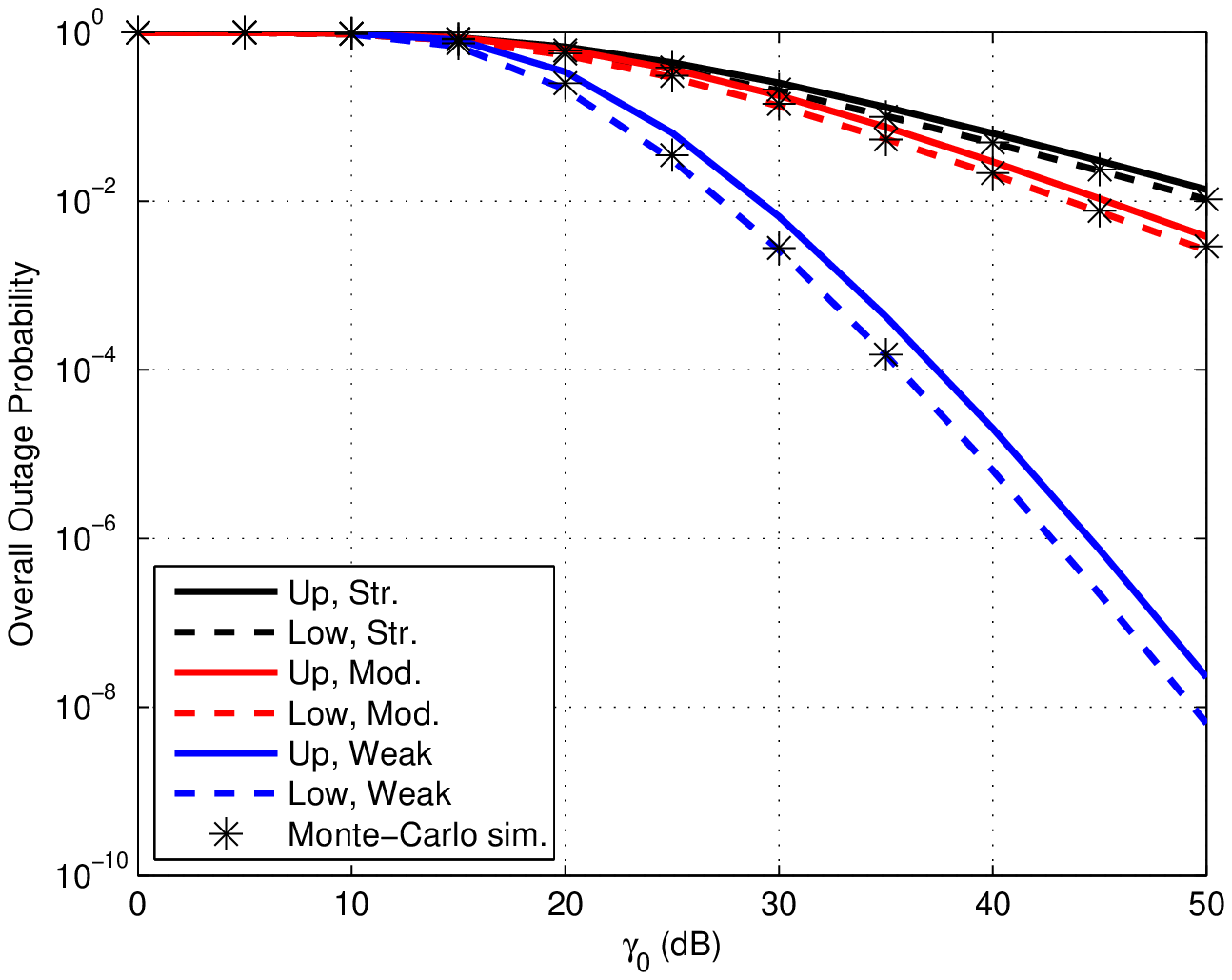}
	\caption{Overall outage probability of a two-way SIM AF relaying system. $J$ is set to 100. $\Gamma_\text{th}$ = 10 dB.}
	\label{fig_outage_10dB}
\end{figure}

\begin{figure}[!t]
	\centering
	\includegraphics[width=3.5in]{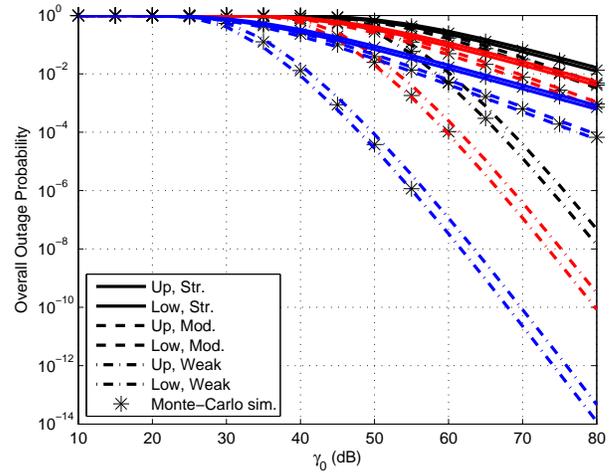}
	\caption{Overall outage probability of a two-way SIM AF relaying system with pointing errors. $J$ is set to 100. $\Gamma_\text{th}$ = 0 dB. $\sigma_s/r$ = 0.1. Black color: $w_z/r$ = 15, Red color: $w_z/r$ = 10, Blue color: $w_z/r$ = 5.}
	\label{fig_outage_withPointError}
\end{figure}

\begin{figure}[!t]
	\centering
	\includegraphics[width=3.5in]{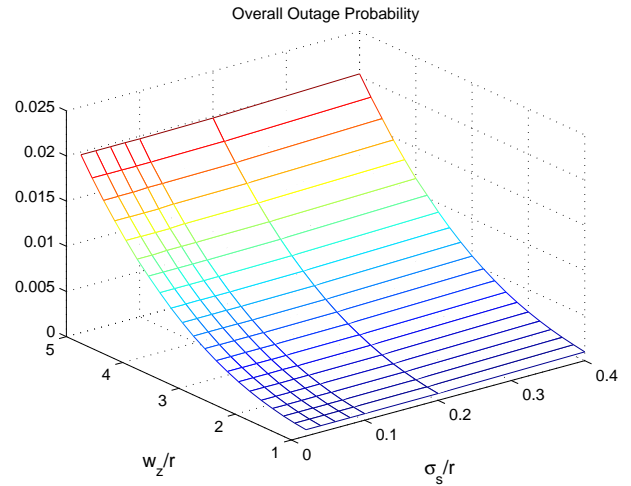}
	\caption{Overall outage probability for the different normalized beamwidth and normalized jitter. $\Gamma_\text{th}$ = 0 dB. $\gamma_0$ = 60 dB. Upper bound for the Strong regime.}
	\label{figOut3D}
\end{figure}

\begin{figure}[!t]
	\centering
	\includegraphics[width=3.5in]{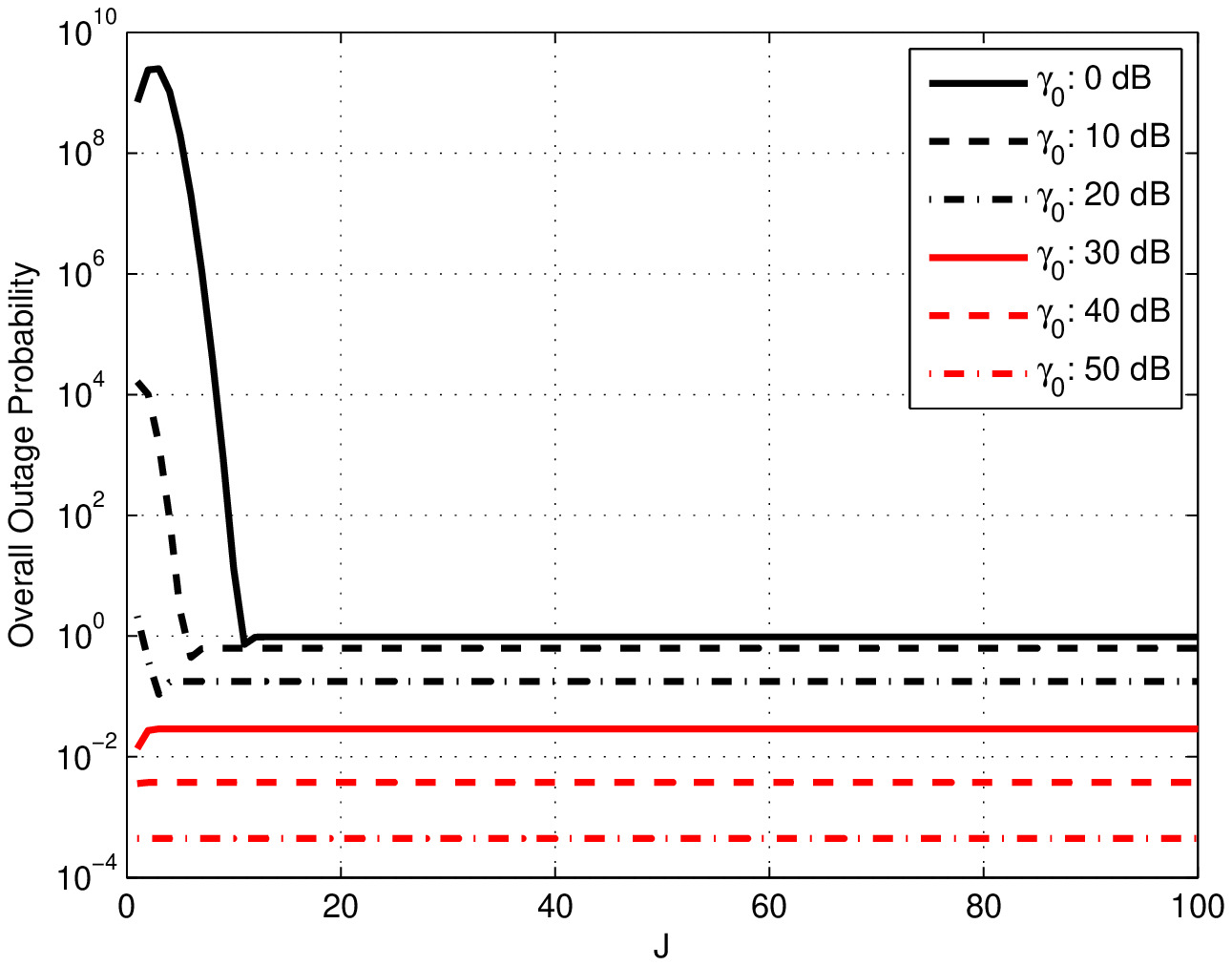}
	\caption{Convergence characteristics of the overall outage probability with respect to $J$ without pointing errors. $\Gamma_\text{th}$ = 0 dB. Upper bound for the Moderate regime.}
	\label{fig_outage_converge}
\end{figure}

According to Fig. 4, which is given for the threshold SNR of 0 dB, to obtain the overall outage probability of $10^{-6}$, the required SNR values are upper bounded by  99.2 dB,  77.8 dB, and 34.6 dB for the strong, moderate, and weak turbulence regimes. It can also be seen in the figure that to obtain the overall outage probability of $10^{-6}$ the required SNR values are lower bounded by  97.6  dB, 76.2 dB, and 32.6 dB for the strong, moderate, and weak turbulence regimes.

According to Fig. 5, which is given for the threshold SNR of 5 dB, to obtain the overall outage probability of $10^{-6}$, the required SNR values are upper bounded by  104.2 dB, 82.8 dB, and 39.6 dB for the strong, moderate, and weak turbulence regimes. It can also be seen in the figure that to obtain the overall outage probability of $10^{-6}$ the required  SNR values are lower bounded by  102.6 dB, 81.2 dB, and 37.6 dB for the strong, moderate, and weak turbulence regimes.

According to Fig. 6, which is given for the threshold SNR of 10 dB, to obtain the overall outage probability of $10^{-6}$, the required SNR values are upper bounded by  109.3 dB, 87.9 dB, and 44.6 dB for the strong, moderate, and weak turbulence regimes. It can also be seen in the figure that to obtain the overall outage probability of $10^{-6}$, the required SNR values are lower bounded by  107.6 dB, 86.2 dB, and 42.6 dB for the strong, moderate, and weak turbulence regimes.

It can be clearly seen in Figs. 4, 5, and 6 that the derived upper and lower bounds, having about 1.6 $\sim$ 2.0 dB gaps, are very tight for the various atmospheric turbulence regimes and SNR scenarios. We also present the Monte-Carlo simulation results of the exact outage probability to evaluate the tightness of the upper and lower bounds of the analyzed outage probability, given in ($\ref{eq:eq_21}$) and ($\ref{eq:eq_26}$). 

\begin{figure}[!t]
	\centering
	\includegraphics[width=3.5in]{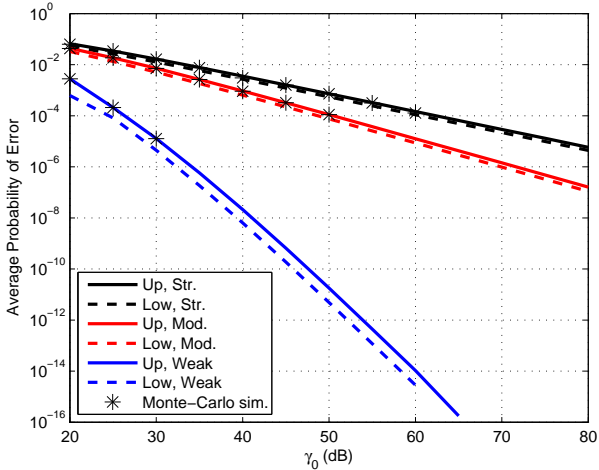}
	\caption{Average probability of error of a two-way AF relaying system for SIM BPSK. $J$ is set to 100.}
	\label{fig_avgBER_SIM}
\end{figure}

\begin{figure}[!t]
	\centering
	\includegraphics[width=3.5in]{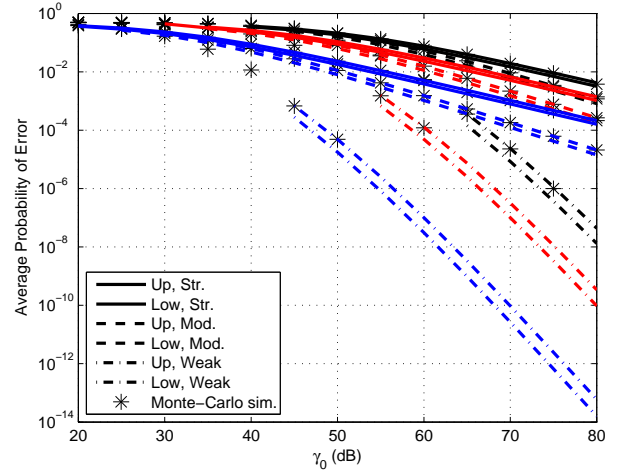}
	\caption{Average probability of error of a two-way AF relaying system for SIM BPSK with pointing errors. $J$ is set to 100. $\sigma_s/r$ = 0.1. Black color: $w_z/r$ = 15, Red color: $w_z/r$ = 10, Blue color: $w_z/r$ = 5.}
	\label{fig_avgBER_SIM_withPointError}
\end{figure}

\begin{figure}[!t]
	\centering
	\includegraphics[width=3.5in]{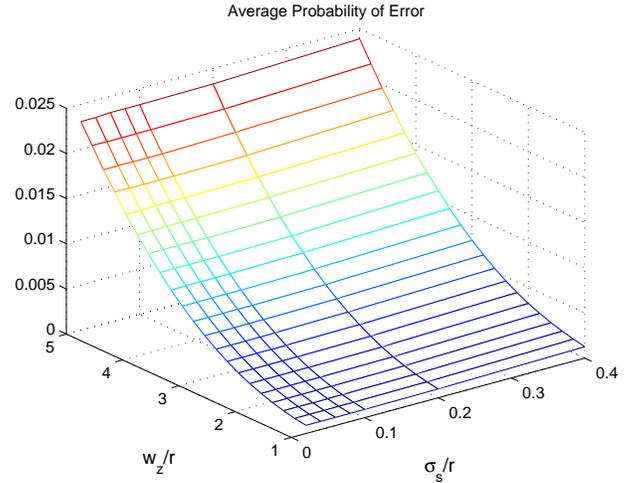}
	\caption{Average probability of error for the different normalized beamwidth and normalized jitter. $\gamma_0$ = 50 dB. Upper bound for the Strong regime.}
	\label{figPe3D}
\end{figure}

\begin{figure}[!t]
	\centering
	\includegraphics[width=3.5in]{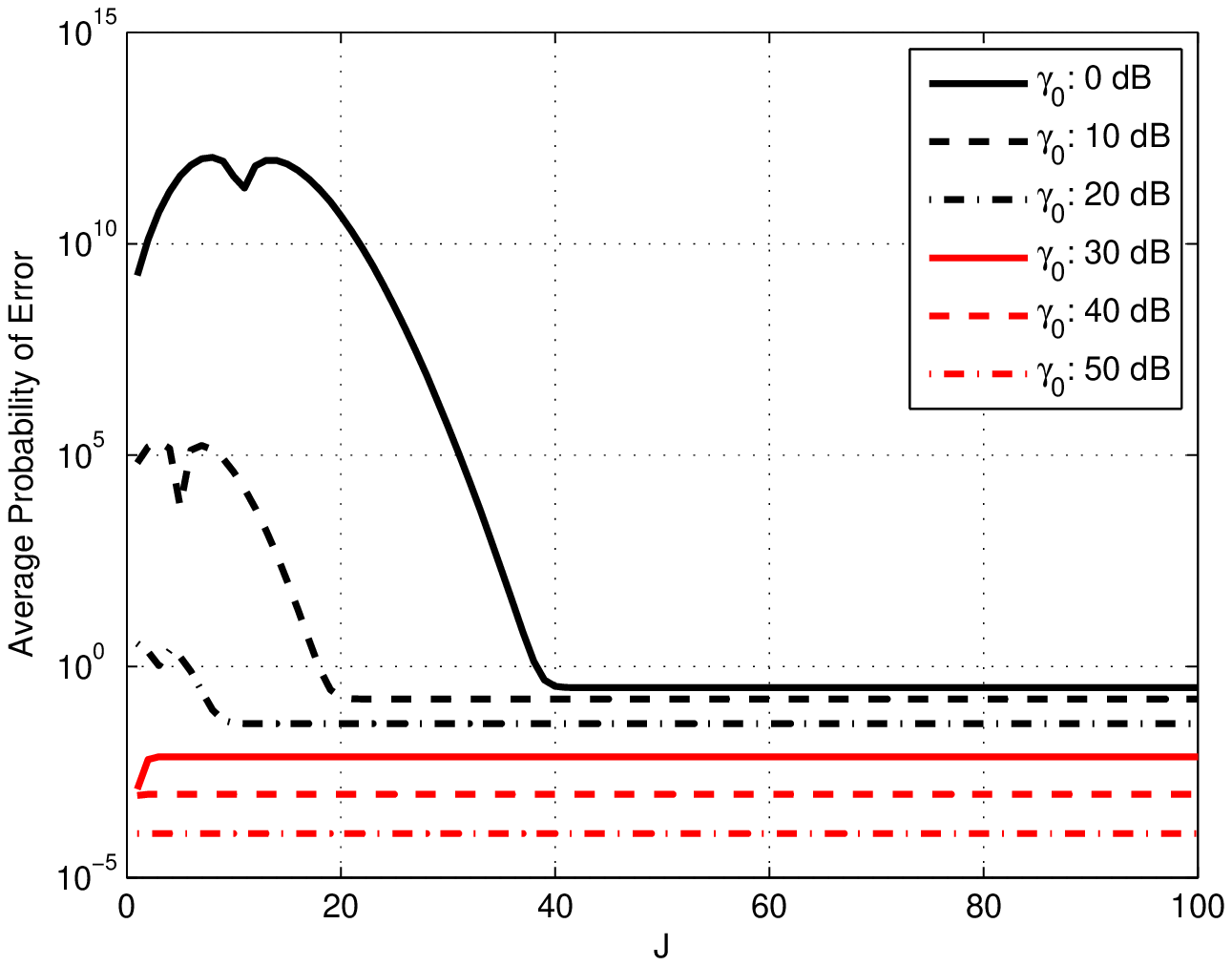}
	\caption{Convergence characteristics of the average probability of error with respect to $J$ without pointing errors. Upper bound for the Moderate regime.}
	\label{fig_avgBER_SIM_converge}
\end{figure}

Figs. 7 and 8 present the overall outage probability performances of the two-way subcarrier intensity-modulated AF relaying system considering the pointing error effects. Fig. 7 shows the overall outage probability performances for various values of the normalized beamwidth $w_z/r = 5, 10, 15$. The threshold SNR is set to 0 dB. For the pointing errors, the normalized jitter $\sigma_s/r$ is set to 0.1. Fig. 8 depicts the upper bound performance of the overall outage probability for the strong regime varying the normalized beamwidths and normalized jitters at the same time. The threshold SNR is set to 0 dB and the SNR is set to 60 dB. It is clearly seen in the figures that a better overall outage probability performance is achieved by using a narrow beamwidth and a small jitter.

Fig. 9 illustrates the convergence performance of the derived overall outage probability with respect to the varying number of power series $J$. The threshold SNR $\Gamma_\text{th}$ is set to 0 dB. Here, the upper bound of the outage probability is considered for the moderate regime. It is observed in the figure that the outage probability converges only with $J=2$ in the high SNR of 50 dB. 

Fig. 10 shows the average probability of error performances of the two-way subcarrier intensity-modulated AF relaying system with respect to the SNR in the absence of turbulence and pointing errors. In this figure, for simplicity's sake, we do not consider pointing errors. As described in Section IV, since the average probability of errors are derived based on the CDF functions from the outage probability's upper and lower bounds, their performances also show the upper and lower bounds. As can be seen in the figure, to obtain the average probability of error of $10^{-6}$ for the BPSK subcarrier intensity modulation scheme, the required SNR values are bounded by 89.5 $\sim$ 91.3 dB, 70.0 $\sim$ 71.8 dB, and 32.3 $\sim$ 34.1 dB for the strong, moderate, and weak turbulence regimes. It can also be clearly seen in the figure that the derived upper and lower bounds have about 1.8 dB gaps for the various atmospheric turbulence regimes and SNR scenarios, which is very tight and similar to the results of the outage probabilities. It should be noted that this two-way subcarrier intensity-modulated AF relaying system using BPSK modulation could be used for practical applications in the case of a weak turbulence regime, in which the required SNR is about 30 dB to obtain the average symbol error probability of $10^{-6}$.

Figs. 11 and 12 present the average probability of error performances of the two-way subcarrier intensity-modulated AF relaying system considering the pointing error effects. Fig.~11 shows the average probability of error performances for various values of the normalized beamwidth $w_z/r = 5, 10, 15$. For the pointing errors, the normalized jitter $\sigma_s/r$ is set to 0.1. Fig.~12 depicts the upper bound performance of the average probability of error for the strong regime varying the normalized beamwidths and normalized jitters at the same time. The SNR is set to 50 dB. It is clearly seen in the figures that a better average probability of error performance is achieved by using a narrow beamwidth and a small jitter.

Fig.~13 shows the convergence performance of the derived average probability of error with respect to the varying number of power series $J$. The pointing error was not considered. Here the upper bound of the average probability of error is considered for the moderate regime. It is observed in the figure that the series term $J=2$ is enough to make the average probability of error converge in the high SNR of 50 dB.

\section{Conclusion} \label{sec:Conclusion}
In this paper, we derived the upper and lower bounds of the overall outage probability and average error probability of a two-way subcarrier intensity-modulated AF relaying system over FSO channels considering attenuations caused by atmospheric turbulence and geometric spread and pointing errors at the same time. The derived performances are based on generalized infinite power series expressions to model the Gamma-Gamma fading distribution of the FSO channel. According to the analysis results, the overall outage probability and average probability of error have only about 2.0 dB and 1.8 dB gaps between the upper and lower bounds. It is noted that this two-way subcarrier intensity-modulated AF relaying system using a BPSK modulation could be used for practical applications in the case of a weak turbulence regime in which the required SNR is about 30 dB to obtain the average probability of error $10^{-6}$. It is also noted that the performance of the two-way subcarrier intensity-modulated AF relaying system is clearly degraded with the pointing errors. We believe that the proposed system could be combined with 5G technologies~\cite{Chae:15a,Chae:15b,Chae:16,Chae:17} and leave this for our future work.

Future work includes

\section*{Appendix A} \label{app:A}
If we substitute ($\ref{eq:eq_22}$) and ($\ref{eq:eq_27}$) into ($\ref{eq:eq_31}$), the average probability of errors corresponding to the upper and lower bounds of the overall outage probability can be expressed, respectively, as

\begin{equation} \label{eq:eq_32}
	P_{U}(e)=\frac{1}{\sqrt{2\pi}} \int_{0}^{\infty}F_{\Gamma}^{U}\left(\frac{x}{\delta}\right)e^{-\frac{x}{2}}\frac{dx}{2\sqrt{x}},
\end{equation}

\begin{equation} \label{eq:eq_33}
	P_{L}(e)=\frac{1}{\sqrt{2\pi}} \int_{0}^{\infty}F_{\Gamma}^{L}\left(\frac{x}{\delta}\right)e^{-\frac{x}{2}}\frac{dx}{2\sqrt{x}}.
\end{equation}
Using (3.361-2) of \cite{Gradshteyn:07},
\begin{equation} \label{eq:eq_34}
	\int_{0}^{\infty}\frac{e^{-qx}}{\sqrt{x}}dx = \sqrt{\frac{\pi}{q}},
\end{equation}
the average probability of errors corresponding to the upper and lower bounds of the overall outage probability can be further expressed as
\begin{equation} \label{eq:eq_35}
	P_{U}(e)=\frac{1}{2\sqrt{2\pi}} \Bigg\{\sqrt{2\pi} - A\left(\frac{\eta^2 \xi^2 \gamma_{0}\delta}{3},\frac{\eta^2 \xi^2 \gamma_{0}\delta}{3}\right)\Bigg\},
\end{equation}
\begin{multline} \label{eq:eq_36}
	P_{L}(e)=\frac{1}{2\sqrt{2\pi}} \Bigg\{\sqrt{2\pi} - A\left(\frac{\eta^2 \xi^2 \gamma_{0}\delta}{2},\frac{\eta^2 \xi^2 \gamma_{0}\delta}{3}\right) \\
	- A\left(\frac{\eta^2 \xi^2 \gamma_{0}\delta}{3},\frac{\eta^2 \xi^2 \gamma_{0}\delta}{2}\right) + A\left(\frac{\eta^2 \xi^2 \gamma_{0}\delta}{3},\frac{\eta^2 \xi^2 \gamma_{0}\delta}{3}\right)  \Bigg\},
\end{multline}
where,
\begin{equation} \label{eq:eq_37}
	A(u,v) = \int_{0}^{\infty} \Psi \left(\frac{u}{x},\frac{v}{x} \right) \frac{e^{-\frac{x}{2}}}{\sqrt{x}} dx.
\end{equation}
If we substitute ($\ref{eq:eq_14}$) into ($\ref{eq:eq_37}$), ($\ref{eq:eq_37}$) can be given by
\begin{equation} \label{eq:eq_38}
	A(u,v) = \sqrt{2\pi} - A_{1}^{J}(u) - A_{2}^{J}(v) + A_{3}^{J}(u,v)
\end{equation}
where,
\begin{equation} \label{eq:eq_39}
	A_{1}^{J}(u) = \int_{0}^{\infty} F_{I_{2}} \left(\sqrt{\frac{x}{u}} \right) \frac{e^{-\frac{x}{2}}}{\sqrt{x}} dx,
\end{equation}

\begin{equation} \label{eq:eq_40}
	A_{2}^{J}(v) = \int_{0}^{\infty} F_{I_{1}} \left(\sqrt{\frac{x}{v}} \right) \frac{e^{-\frac{x}{2}}}{\sqrt{x}} dx,
\end{equation}

\begin{equation} \label{eq:eq_41}
	A_{3}^{J}(u,v) = \int_{0}^{\infty}  F_{I_{2}} \left(\sqrt{\frac{x}{u}} \right) F_{I_{1}} \left(\sqrt{\frac{x}{v}} \right)
	\frac{e^{-\frac{x}{2}}}{\sqrt{x}} dx.
\end{equation}
If we substitute ($\ref{eq:eq_cdfcombinedfinal}$) into ($\ref{eq:eq_39}$) and ($\ref{eq:eq_40}$), $A_{1}^{J}(u)$ and $A_{2}^{J}(v)$ can be derived as, respectively,
\begin{multline} \label{eq:eq_42}
	A_{1}^{J}(u) = \frac{\gamma^2}{A_0} \lim_{J \to \infty} \sum_{j=0}^J \frac{a_{j}(\alpha,\beta)u^{-\frac{j+\beta}{2}}}{(j+\beta)(\gamma^2-j-\beta)A_0^{j+\beta-1}} \\
	\times \int_{0}^{\infty} x^{\frac{j+\beta+1}{2}-1} e^{-\frac{x}{2}} dx\\
	+ \frac{\gamma^2}{A_0} \lim_{J \to \infty} \sum_{j=0}^J \frac{a_{j}(\beta,\alpha)u^{-\frac{j+\alpha}{2}}}{(j+\alpha)(\gamma^2-j-\alpha)A_0^{j+\alpha-1}} \\
	\times \int_{0}^{\infty} x^{\frac{j+\alpha+1}{2}-1} e^{-\frac{x}{2}} dx,~~~~~~~~~~~~~
\end{multline}
\begin{multline} \label{eq:eq_43}
	A_{2}^{J}(v) = \frac{\gamma^2}{A_0} \lim_{J \to \infty} \sum_{j=0}^J \frac{a_{j}(\alpha,\beta)v^{-\frac{j+\beta}{2}}}{(j+\beta)(\gamma^2-j-\beta)A_0^{j+\beta-1}} \\
	\times \int_{0}^{\infty} x^{\frac{j+\beta+1}{2}-1} e^{-\frac{x}{2}} dx\\
	+ \frac{\gamma^2}{A_0} \lim_{J \to \infty} \sum_{j=0}^J \frac{a_{j}(\beta,\alpha)v^{-\frac{j+\alpha}{2}}}{(j+\alpha)(\gamma^2-j-\alpha)A_0^{j+\alpha-1}} \\
	\times \int_{0}^{\infty} x^{\frac{j+\alpha+1}{2}-1} e^{-\frac{x}{2}} dx.~~~~~~~~~~~~~
\end{multline}
If we apply the gamma function [35, eq. (3.381-4)] to ($\ref{eq:eq_42}$) and ($\ref{eq:eq_43}$)
\begin{equation} \label{eq:eq_44}
	\int_{0}^{\infty} x^{n-1} e^{-mx} dx = \frac{1}{m^n} \Gamma(n),
\end{equation}
then, $A_{1}(u)$ and $A_{2}(v)$ can be finally obtained as given by ($\ref{eq:eq_A1final}$) and ($\ref{eq:eq_A2final}$).
If we substitute ($\ref{eq:eq_cdfcombinedfinal}$) into the product of CDFs in ($\ref{eq:eq_41}$), it can be expressed as
\begin{multline} \label{eq:eq_47}
	F_{I_{2}}\left(\sqrt{\frac{x}{u}} \right) F_{I_{1}} \left(\sqrt{\frac{x}{v}} \right)\\
	= \left(\frac{\gamma^2}{A_0}\right)^2 \lim_{J \to \infty}
	\sum_{j=0}^J \frac{a_{j}(\alpha,\beta)u^{-\frac{j+\beta}{2}}}{(j+\beta)(j+\beta-\gamma^2)A_0^{j+\beta-1}} x^{j/2}\\
	\times \sum_{j=0}^J \frac{a_{j}(\alpha,\beta)v^{-\frac{j+\beta}{2}}}{(j+\beta)(j+\beta-\gamma^2)A_0^{j+\beta-1}} x^{j/2} x^{\beta}\\
	+ \left(\frac{\gamma^2}{A_0}\right)^2 \lim_{J \to \infty}
	\sum_{j=0}^J \frac{a_{j}(\alpha,\beta)u^{-\frac{j+\beta}{2}}}{(j+\beta)(j+\beta-\gamma^2)A_0^{j+\beta-1}} x^{j/2}\\
	\times \sum_{j=0}^J \frac{a_{j}(\beta,\alpha)v^{-\frac{j+\alpha}{2}}}{(j+\alpha)(j+\alpha-\gamma^2)A_0^{j+\alpha-1}} x^{j/2} x^{\frac{\alpha+\beta}{2}}\\
	+ \left(\frac{\gamma^2}{A_0}\right)^2 \lim_{J \to \infty}
	\sum_{j=0}^J \frac{a_{j}(\beta,\alpha)u^{-\frac{j+\alpha}{2}}}{(j+\alpha)(j+\alpha-\gamma^2)A_0^{j+\alpha-1}} x^{j/2} \\
	\times \sum_{j=0}^J \frac{a_{j}(\alpha,\beta)v^{-\frac{j+\beta}{2}}}{(j+\beta)(j+\beta-\gamma^2)A_0^{j+\beta-1}} x^{j/2}x^{\frac{\alpha+\beta}{2}}\\
	\end{multline}
\begin{multline}
	+ \left(\frac{\gamma^2}{A_0}\right)^2 \lim_{J \to \infty}
	\sum_{j=0}^J \frac{a_{j}(\beta,\alpha)u^{-\frac{j+\alpha}{2}}}{(j+\alpha)(j+\alpha-\gamma^2)A_0^{j+\alpha-1}} x^{j/2} \\
	\times \sum_{j=0}^J \frac{a_{j}(\beta,\alpha)v^{-\frac{j+\alpha}{2}}}{(j+\alpha)(j+\alpha-\gamma^2)A_0^{j+\alpha-1}} x^{j/2}x^{\alpha}.\nonumber
\end{multline}
If we apply the multiplication of power series [35, eq. (0.316)]
\begin{equation} \label{eq:eq_48}
	\sum_{j=0}^\infty b_{j}y^{j} \sum_{j=0}^\infty d_{j}y^{j} = \sum_{j=0}^\infty c_{j}y^{j}, c_{j}=\sum_{k=0}^j b_{k}d_{j-k},
\end{equation}
the product of CDFs can be expressed as
\begin{multline} \label{eq:eq_cdfProduct}
	F_{I_{2}}\left(\sqrt{\frac{x}{u}} \right) F_{I_{1}} \left(\sqrt{\frac{x}{v}} \right)\\
	= \left(\frac{\gamma^2}{A_0}\right)^2 \lim_{J \to \infty} \sum_{j=0}^J c_{j}^{(1p)}(u,v) x^{\frac{j}{2}+\beta} ~~\\
	+ \left(\frac{\gamma^2}{A_0}\right)^2 \lim_{J \to \infty} \sum_{j=0}^J c_{j}^{(2p)}(u,v) x^{\frac{j}{2}+\frac{\alpha+\beta}{2}}\\
	+ \left(\frac{\gamma^2}{A_0}\right)^2 \lim_{J \to \infty} \sum_{j=0}^J c_{j}^{(3p)}(u,v) x^{\frac{j}{2}+\frac{\alpha+\beta}{2}} \\
	+ \left(\frac{\gamma^2}{A_0}\right)^2 \lim_{J \to \infty} \sum_{j=0}^J c_{j}^{(4p)}(u,v) x^{\frac{j}{2}+\alpha}~~~~~~~~
\end{multline}
where,
\begin{multline} \label{eq:eq_cjnpfunc}
	c_{j}^{(1p)}(u,v) = \sum_{k=0}^j \frac{a_{k}(\alpha,\beta)u^{-\frac{k+\beta}{2}}}{(k+\beta)(k+\beta-\gamma^2)A_0^{k+\beta-1}} \\ \times \frac{a_{j-k}(\alpha,\beta)v^{-\frac{j-k+\beta}{2}}}{(j-k+\beta)(j-k+\beta-\gamma^2)A_0^{j-k+\beta-1}} \\
	c_{j}^{(2p)}(u,v) = \sum_{k=0}^j \frac{a_{k}(\alpha,\beta)u^{-\frac{k+\beta}{2}}}{(k+\beta)(k+\beta-\gamma^2)A_0^{k+\beta-1}} \\ \times \frac{a_{j-k}(\beta,\alpha)v^{-\frac{j-k+\alpha}{2}}}{(j-k+\alpha)(j-k+\alpha-\gamma^2)A_0^{j-k+\alpha-1}} \\
	c_{j}^{(3p)}(u,v) = \sum_{k=0}^j \frac{a_{k}(\beta,\alpha)u^{-\frac{k+\alpha}{2}}}{(k+\alpha)(k+\alpha-\gamma^2)A_0^{k+\alpha-1}} \\ \times
	\frac{a_{j-k}(\alpha,\beta)v^{-\frac{j-k+\beta}{2}}}{(j-k+\beta)(j-k+\beta-\gamma^2)A_0^{j-k+\beta-1}} \\
	c_{j}^{(4p)}(u,v) = \sum_{k=0}^j \frac{a_{k}(\beta,\alpha)u^{-\frac{k+\alpha}{2}}}{(k+\alpha)(k+\alpha-\gamma^2)A_0^{k+\alpha-1}} \\ \times
	\frac{a_{j-k}(\beta,\alpha)v^{-\frac{j-k+\alpha}{2}}}{(j-k+\alpha)(j-k+\alpha-\gamma^2)A_0^{j-k+\alpha-1}}. \\
\end{multline}
If we substitute the product of CDFs ($\ref{eq:eq_cdfProduct}$) into ($\ref{eq:eq_41}$), $A_{3}(u,v)$ can be derived as
\begin{multline} \label{eq:eq_A3semifinal}
	A_{3}^{J}(u,v)\\
	= \left(\frac{\gamma^2}{A_0}\right)^2 \lim_{J \to \infty} \sum_{j=0}^J c_{j}^{(1p)}(u,v) \int_{0}^{\infty}  x^{\frac{j+2\beta+1}{2}-1} e^{-\frac{x}{2}}dx \\
	+ \left(\frac{\gamma^2}{A_0}\right)^2 \lim_{J \to \infty} \sum_{j=0}^J c_{j}^{(2p)}(u,v) \int_{0}^{\infty}  x^{\frac{j+\alpha+\beta+1}{2}-1} e^{-\frac{x}{2}}dx\\
	+ \left(\frac{\gamma^2}{A_0}\right)^2 \lim_{J \to \infty} \sum_{j=0}^J c_{j}^{(3p)}(u,v) \int_{0}^{\infty}  x^{\frac{j+\alpha+\beta+1}{2}-1} e^{-\frac{x}{2}}dx\\
	+ \left(\frac{\gamma^2}{A_0}\right)^2 \lim_{J \to \infty} \sum_{j=0}^J c_{j}^{(4p)}(u,v) \int_{0}^{\infty}  x^{\frac{j+2\alpha+1}{2}-1} e^{-\frac{x}{2}}dx.~
\end{multline}
If we apply the gamma function [35, eq. (3.381-4)] to ($\ref{eq:eq_A3semifinal}$), $A_{3}(u,v)$  can be finally obtained as given by ($\ref{eq:eq_A3final}$).

\vspace{10pt}
\bibliographystyle{IEEE}

\epsfysize=3.2cm
\begin{biography}{Jaedon Park} received the B.S. degree in Electronics Engineering from Hanyang University, Seoul, Korea in 2000, and the M.S. and Ph.D. degrees in the School of Electrical Engineering from the Korea Advanced Institute of Science and Technology (KAIST), Daejeon, Korea in 2002 and 2016, respectively. Currently, he is a Senior Researcher in the Agency for Defense Development (ADD), Daejeon, Korea. His research interests are in MIMO systems, relay systems, and FSO systems.
\end{biography}
\epsfysize=3.2cm
\begin{biography}{Chan-Byoung Chae}
	(S'06 - M'09 - SM'12)  is the Underwood Distinguished Professor, in the School of Integrated Technology, College of Engineering, Yonsei University, Korea. He was a Member of Technical Staff (Research Scientist) at Bell Laboratories, Alcatel-Lucent, Murray Hill, NJ, USA from 2009 to 2011. Before joining Bell Laboratories, he was with the School of Engineering and Applied Sciences at Harvard University, Cambridge, MA, USA as a Post-Doctoral Research Fellow. He received the Ph.D. degree in Electrical and Computer Engineering from The University of Texas (UT), Austin, TX, USA in 2008, where he was a member of the Wireless Networking and Communications Group (WNCG). 	
	Prior to joining UT, he was a Research Engineer at the Telecommunications R\&D Center, Samsung Electronics, Suwon, Korea, from 2001 to 2005.  While having worked at Samsung, he participated in the IEEE 802.16e (mobile WiMAX) standardization, where he made several contributions and filed a number of related patents from 2004 to 2005. His current research interests include capacity analysis and interference management in energy-efficient wireless mobile networks and nano (molecular) communications. He has served/serves as an Editor for the \textsc{IEEE Comm. Mag.}, the \textsc{IEEE Wireless Comm. Letters}, the \textsc{IEEE Trans. on Wireless Comm.}, the \textsc{IEEE Trans. on Molecular, Biological, Multi-scale Comm.}, the \textsc{IEEE Trans. on Smart Grid}, and the \textsc{IEEE/KICS Jour. Comm. Nets}. He was a Guest Editor for the \textsc{IEEE Jour. Sel. Areas in Comm.} (special issue on molecular, biological, and multi-scale comm.). He is an IEEE Senior Member.
	
	Dr. Chae was the recipient/co-recipient of the Outstanding Teaching Award (2017) from Yonsei University, the Underwood Distinguished Professor Award (2016), the Yonam Research Award from LG Yonam Foundation (2016), the Best Young Professor Award from the College of Engineering, Yonsei University (2015), the IEEE INFOCOM Best Demo Award (2015), the IEIE/IEEE Joint Award for Young IT Engineer of the Year (2014), the Haedong Young Scholar Award (2013), the IEEE Signal Processing Magazine Best Paper Award (2013), the IEEE ComSoc AP Outstanding Young Researcher Award (2012), the IEEE VTS Dan. E. Noble Fellowship Award (2008), the Gold Prize (1st) in the 14th/19th Humantech Paper Contests, and the KSEA-KUSCO scholarship (2007). He also received the Korea Government Fellowship (KOSEF) during his Ph.D. studies.\end{biography}

\epsfysize=3.2cm
\begin{biography}{Giwan Yoon} received the B.S. degree from Seoul National University (SNU), Seoul, Korea, in 1983, the M.S. degree from the Korea Advanced Institute of Science and Technology (KAIST), Seoul, Korea, in 1985, and the Ph.D. degree from the University of Texas at Austin, USA, in 1994. From 1985 to 1990, he was employed as an engineer at LG Group, Seoul, Korea, where he worked for the development of high-speed bipolar transistors. From 1994 to 1997, he was employed as a senior engineer at Digital Equipment Corporation (DEC), MA, USA, where he developed oxynitride gate dielectric CMOS devices. From 1997 to 2009, he was a faculty member of Information and Communications University, Daejeon, Korea, where he developed high-frequency devices for RF and wireless communications. Since 2009, he has been with the KAIST, where he is currently a professor in the School of Electrical Engineering with teaching and research activities in the areas of nano devices and integrated systems, energy generation \& harvesting devices, and flexible/wearable sensing devices for healthcare, IOT and sensor networks applications. 
Dr. Yoon is a member of the IEEE.
\end{biography}

\end{document}